\documentclass{article}

\usepackage{PRIMEarxiv}

\usepackage[utf8]{inputenc}
\usepackage[T1]{fontenc}
\usepackage{hyperref}
\usepackage{url}
\usepackage{booktabs}
\usepackage{amsfonts}
\usepackage{nicefrac}
\usepackage{microtype}
\usepackage{lipsum}
\usepackage{fancyhdr}
\usepackage{graphicx}
\usepackage{multirow}
\graphicspath{{media/}}

\title{EMMR: Emotion-Mediated Multimodal Reasoning for Personality Assessment in Asynchronous Video Interviews}

\author{
Dongsheng~Hu \\
Hangzhou Normal University \\
\texttt{2024112013041@stu.hznu.edu.cn}
\And
Tianyi~Zhang$^{*}$ \\
Southeast University \\
\texttt{t.zhang@seu.edu.cn}
\And
Chuang~Liu \\
Hangzhou Normal University \\
\texttt{liuchuang@hznu.edu.cn}
\AND
Yuan~Zong \\
Southeast University \\
\texttt{xhzongyuan@seu.edu.cn}
\And
Yong~Li \\
Southeast University \\
\texttt{yong.li@seu.edu.cn}
\And
Wenming~Zheng \\
Southeast University \\
\texttt{wenmingzheng@seu.edu.cn}
\AND
XiuXiu~Zhan$^{*}$ \\
Hangzhou Normal University \\
\texttt{zhanxiuxiu@hznu.edu.cn}
}

\begin{document}
\maketitle

\begin{abstract}
Asynchronous Video Interviews (AVIs) have become increasingly popular for personality assessment. Recent large language models (LLMs) have shown potential for personality assessment from transcribed interview responses. However, text-centered methods may overlook non-verbal behavioral cues conveyed through visual and audio modalities, even though such cues are highly relevant to personality assessment. In particular, emotion-related cues provide important social and affective evidence for understanding candidates' behavior related to personality traits. Thus, we propose \textit{EMMR} (Emotion-Mediated Multimodal Reasoning), a two-stage framework for MLLMs-based personality assessment for AVIs. \textit{EMMR} extracts emotion-related cues from multimodal interview data and incorporates them into personality assessment through structured reasoning as auxiliary social and behavioral evidence. Experiments on two AVIs datasets, OPVA and AVI-6, show that \textit{EMMR} improves MAE, MSE, and PCC compared with baselines. Further analysis indicates that semantic descriptions of emotion cues enhance personality assessment, while their quality affects personality assessment reliability. These results suggest that integrating emotion-related cues into multimodal reasoning is a promising direction for more interpretable MLLMs-based personality assessment in AVIs.
\end{abstract}

% keywords can be removed

\section{Introduction}
In recent years, asynchronous video interviews (AVIs) have become a common method in remote recruitment, especially after the COVID-19 pandemic \cite{lukacik2022into},\cite{mehta2020recent}. Since AVIs allow interviewees to record video responses to pre-set questions, they can not only serve as a new form between interviewees and interviewers, but also provide employers with more information about candidates’ behaviors and personality traits.
Many companies use AVIs to assess personality traits, which are regarded as important indicators of a person's suitability for a job and workplace culture \cite{hickman2022automated},\cite{liff2024psychometric}. While recruiters often evaluate personality by inviting human resource experts to watch the videos. Artificial intelligence (AI) tools, especially those based on deep learning methods, are now widely used for automated assessment. Among various AI models, large language models (LLMs) such as GPT-4\cite{achiam2023gpt}, Qwen-2.5\cite{hui2024qwen2}, and Gemma-2\cite{team2024gemma} have shown great potential for personality assessment. Their extensive training on large-scale text data enables them to understand and assess personality traits from transcribed interview responses. Their ability to generate textual rationales may also provide a certain degree of transparency for personality assessment outputs compared with traditional learning models\cite{tan2025prompting}.

Although previous studies have explored the application of LLMs in AVIs-based personality assessment\cite{tan2025prompting,zhang2024can,zhang2026mixture,11481797}, a key limitation remains that most existing methods primarily rely on the transcribed interview responses. Admittedly, textual information is important and remain the most mature modality for LLMs-based processing. However, personality assessment in AVIs are not formed from textual information alone. Brunswik's Lens Model\cite{breil202113} suggests that personality assessment are formed through multiple observable cues. The Realistic Accuracy Model \cite{Realistic} suggests that accurate personality assessment depends on the relevance, availability, detection, and utilization of behavioral cues. Therefore, relying only on textual information may provide an incomplete basis for personality assessment, especially in AVIs.
More specifically, relying only on transcribed responses may lead to misinterpretation when verbal content conflicts with non-verbal behavioral cues \cite{9760455,8999746,9792204}. For example, interviewees may describe themselves as confident and proactive, but their hesitant voice, tense facial expressions, or limited expressiveness may suggest anxiety or low assertiveness. In such cases, text-only methods may overestimate traits such as Extraversion by focusing mainly on the transcribed responses\cite{9760455,koutsoumpis2024beyond}. Therefore, incorporating multimodal information can provide a more complete basis for personality assessment.
 
Recently, multimodal large language models (MLLMs) have shown increasing potential for integrating visual, audio, and textual information within a unified reasoning framework\cite{yin2024survey}. 
Despite their strong multimodal understanding capabilities, existing MLLMs are not specifically designed for personality assessment in AVIs. Directly applying MLLMs to this task remains challenging, because AVIs-based personality assessment requires interpreting multimodal information as personality-related social and behavioral evidence within the interview context \cite{hickman2022automated,suen2024artificial,zhang2025assessing}, rather than merely processing multimodal inputs. Among different types of such evidence, emotion cues are particularly relevant to personality assessment. 
Psychological studies have shown that emotion cues are closely associated with personality traits\cite{9868797,marengo2021meta,sacchi2026understanding,9210819,zhu2023understanding,cai2025mdpe,jie2022impact,11458684}. For example, individuals with higher Extraversion tend to exhibit stronger positive emotional responses\cite{jie2022impact}. 
These findings provide a theoretical basis for considering emotion cues as auxiliary social and behavioral evidence in personality assessment. 
However, emotion cues can be expressed through multiple modalities, including facial expressions, vocal tone, and linguistic content\cite{9969993,zhang2024deep}. This multimodal nature suggests that emotion cues in AVIs should not be treated as isolated labels, but as behavioral evidence distributed across visual, audio, and textual information. 

Motivated by this perspective, we propose \textit{EMMR} (\textit{Emotion-Mediated Multimodal Reasoning}), a two-stage framework for MLLMs-based personality assessment in AVIs. In the first stage, \textit{EMMR} extracts emotion cues from visual, audio, and textual modalities, transforming these heterogeneous multimodal signals into rich, structured semantic descriptions. 
In the second stage, the semantic descriptions of emotion cues are first integrated with interview responses, and then incorporated into MLLMs-based personality assessment through structured reasoning. This design enables the model to jointly reason over candidates' verbal content and nonverbal behavioral cues, providing a unified framework for analyzing the role of multimodal emotion cues in personality assessment in AVIs. The contributions of this study are summarized as follows:

\begin{itemize}
    \item We propose \textit{EMMR} (\textit{Emotion-Mediated Multimodal Reasoning}), a two-stage framework that integrates multimodal semantic descriptions of emotion cues as structured auxiliary evidence to enhance MLLMs-based personality assessment in AVIs.
    
    \item We extensively evaluate \textit{EMMR} on two AVI datasets (OPVA and AVI-6) under zero-shot settings, demonstrating that it significantly outperforms diverse baseline models. By achieving lower MAE, MSE and higher PCC across various HEXACO dimensions, we validate its superior accuracy, stability, and robustness under both rich and sparse context conditions.

    \item We conduct further analyses to examine the role of emotion cues in MLLMs-based personality assessment. Through modality ablation, emotion cue comparison, and error analysis, we provide valuable empirical evidence on the contribution and limitations of emotion cues in MLLMs-based personality assessment.
\end{itemize}

\section{Related Work}

In this section, we first introduce the HEXACO personality model, which is used to model personality traits in our work. Secondly, we review the literature on personality assessment from AVIs, focusing on the techniques employed to quantify personality traits from video data. Afterwards, we review the rapid development of LLMs. In this part, we discuss their advantages and limitations in the context of personality assessment. Finally, we summarize previous work on MLLMs methods and explore the potential strategy to improve the validity of personality assessment.

\subsection{HEXACO personality model}

The HEXACO personality model\cite{ashton2007empirical} is a widely adopted framework for assessing personality traits in AVIs. It comprises six major dimensions: \textbf{Honesty-Humility (H), Emotionality (E), eXtraversion (X), Agreeableness (A), Conscientiousness (C), and Openness to Experience (O)}. Unlike the Big Five model, the HEXACO personality model introduces the Honesty-Humility factor to capture traits such as sincerity, fairness, and modesty\cite{ashton2004six}. These attributes are particularly relevant in moral decision-making and integrity based on workplace behaviors. Each of these six factors is further divided into four facets. This rich structure has demonstrated predictive validity in job-related domains. For example, individuals with high Emotionality may be more sensitive to emotion cues in the workplace, which affects their career choice and job satisfaction\cite{krivoshchekov2024passion}. Thus, HEXACO has been widely used to predict leadership effectiveness, team collaboration compatibility, and employee well-being. Given these advantages, we adopt the HEXACO model to define the ground-truth personality scores for AVIs personality assessment.
%High Extraversion is positively correlated with leadership performance and team communication efficiency in dynamic work environments\cite{a2024promoting}.

\subsection{AVIs-based personality assessment}

Asynchronous video interviews (AVIs) are designed to capture candidates' responses to work-related questions and provide insights into candidates' personality traits or other desired outcome variables\cite{hickman2022automated,liff2024psychometric,zhang2025assessing}. Unlike personality cues derived from online videos or everyday conversations (e.g., ChaLearn \cite{ponce2016chalearn} or PersonalityEvd \cite{sun2024revealing}), AVIs use questions that are often carefully crafted by psychologists to elicit responses reflecting different personality traits\cite{zhang2025assessing}. This structured design makes sure that candidates’ responses truly reflect their underlying personality traits, as the questions are explicitly intended to activate them. 

Recent studies on AVIs-based personality assessment generally categorize existing methods into two groups, verbal methods and non-verbal methods. Verbal methods perform personality assessment by analyzing the transcribed responses of candidates. 
For example, Marouf et al.\cite{9003524} conduct a comparative study of feature selection algorithms for social media based computational personality assessment. Pavan Kumar and Gavrilova\cite{9531972} propose a contextual language embedding approach for latent personality assessment from social network activity.

Non-verbal methods perform personality assessment using multimodal information from visual and audio modalities, without relying on transcribed responses \cite{9760455,hickman2022automated}. For example, Sun et al.\cite{9760455} proposed a modality disentangling framework for personality assessment in short video, which models visual, acoustic, and textual cues with temporal alignment and shared private representation learning to capture common and modality specific personality assessment.
Koutsoumpis et al. \cite{koutsoumpis2024beyond} proposed a machine learning-based psychometric evaluation framework for AVIs, integrating multimodal behavioral cues and psychometric analysis for personality assessment.

Verbal and non-verbal methods capture distinct dimensions of candidate behavior in AVIs. While verbal methods analyze transcribed responses to yield valuable semantic insights, they frequently overlook facial and vocal nuances. Conversely, non-verbal methods evaluate behavioral cues but often lack the contextual depth of the actual responses. These inherent limitations indicate that personality assessments in AVIs should integrate textual, visual, and acoustic data.

\subsection{LLMs-based personality assessment}

LLMs are advanced neural architectures trained on massive text corpora, enabling them to capture complex linguistic patterns and semantic representations. Different from traditional approaches that depend on handcrafted or supervised training on large labeled datasets, LLMs demonstrate remarkable capabilities such as zero-shot or few-shot learning, lingual generalization, and adaptability to diverse textual contexts.

Based on these advantages, many researchers have developed automated personality assessment algorithms using LLMs.
Zhang et al. \cite{zhang2024can} comprehensively evaluated the validity, reliability, fairness, and scoring patterns of GPT-3.5 and GPT-4 in personality assessment. Their findings showed that the zero-shot validity of LLMs and the score explanations generated by these models are generally consistent with psychological theories.
Tan et al.\cite{tan2025prompting} propose PICEPR, which integrates a "Prompting-in-a-Series" algorithm, content and embedding pipelines, and modularised decoder-only LLMs, functioning as a personality feature extractor and rich-content generator that achieves a new state-of-the-art performance with a 5–15\% improvement in personality assessment.
Zhang et al.\cite{zhang2026mixture} proposed MoE-Personality for text based personality assessment from AVIs, which uses multiple annotator LLMs experts and an aggregator LLMs expert to produce fine grained trait scores and reduce positivity bias.

Although LLMs-based methods have been used for personality assessment, most of them still rely mainly on transcribed responses, linguistic patterns, and generated explanations. This makes these methods easy to apply and relatively easy to interpret, but it also narrows the evidence used for personality assessment. Facial expressions, vocal characteristics, and emotion cues may also provide useful information\cite{hickman2022automated,suen2024artificial}, but they are often not explicitly modeled in LLMs-based methods when applied to AVIs. When personality assessment relies solely on textual information, discrepancies may arise between textual content and candidates' nonverbal signals, which may introduce modality conflict\cite{9760455,8999746,9792204}. Therefore, existing studies have demonstrated the feasibility of text driven personality assessment, but the use of visual, audio, and emotion cues in MLLMs-based personality assessment remains underexplored.

\subsection{Multimodal integration in LLMs-based personality assessment}

Text-only personality assessment is limited because it cannot fully use non-verbal cues in AVIs. Thus,  recent studies have begun to incorporate LLMs into multimodal personality assessment, using them to enhance semantic understanding, generate psychological descriptions, or support cross-modal reasoning.
For example,Yang et al.\cite{yang2025enhancing} proposed an LLMs augmented hierarchical fusion framework for multimodal personality assessment in AVIs, which uses LLMs generated psychological descriptions and multi path fusion to enhance semantic cues and cross modal interactions.
Kassab et al.\cite{kassab2026multi} proposed an interpretable multimodal framework for candidate interview assessment, which combines non-verbal behavior, LLMs based verbal analysis, and Big Five traits into theory based constructs to produce a transparent Top Potential Score.

These studies show that LLMs can enhance multimodal personality assessment by improving semantic representation and supporting interpretable reasoning. However, they still rely on additional feature extraction or fusion modules to combine verbal and nonverbal information. Recent multimodal large language models (MLLMs) provide a more direct way to process visual, audio, and textual inputs within a unified framework \cite{yin2024survey,11194051}. This capability makes MLLMs inherently promising for AVIs-based personality assessment, where candidates' responses naturally contain rich multimodal information.

Despite this architectural advantage, the full potential of MLLMs in this domain requires more than simply inputting raw multimodal data. The visual, audio, and textual information must be explicitly interpreted and organized as personality-related social and behavioral evidence. Among these, emotion cues are particularly relevant as they can be consistently observed across all modalities \cite{marengo2021meta,sacchi2026understanding}. However, current research has not sufficiently examined how such structured emotion cues should be represented and incorporated to guide MLLMs-based trait-level reasoning.

\begin{figure*}[htbp]
\centering
\includegraphics[width=1\textwidth]{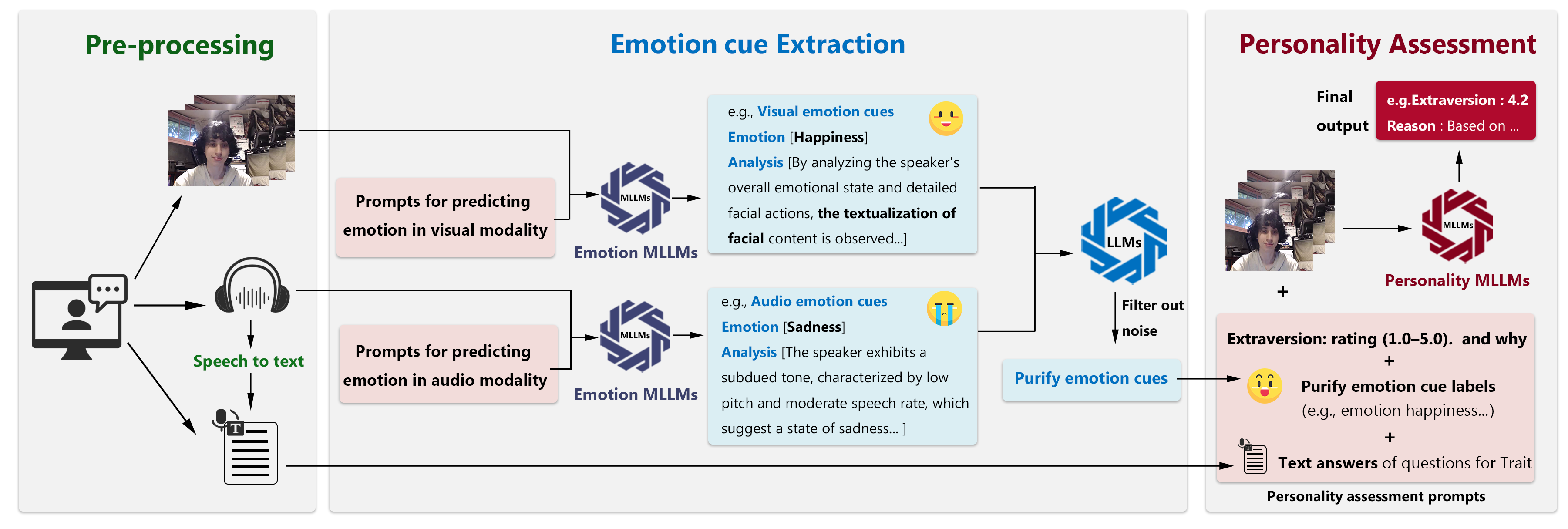}
\caption{The framework of \textit{EMMR}: emotion cues are first extracted from multimodal inputs through the emotion cue extraction module. These cues are then integrated with the interview responses in the emotion cue fusion module. Finally, the combined information is incorporated into MLLMs-based reasoning for personality assessment, enabling joint reasoning over verbal content and non-verbal behavioral cues.}
\label{fig.full_framework}
\end{figure*}

\section{Methodology}

In this section, we introduce \textit{EMMR}, a two-stage framework for MLLMs-based personality assessment in AVIs. The framework leverages multimodal interview data, including visual, audio, and textual information, and incorporates emotion cues as auxiliary social and behavioral evidence for trait-level reasoning.
The proposed framework consists of two modules. \textbf{(1) Emotion cue extraction module.} This module processes visual and audio modalities to extract modality-specific emotion cues, including facial expressions and facial dynamics from visual inputs, as well as prosodic features such as tone fluctuations and speech rhythm from audio inputs. To support reasoning within the language space, these multimodal emotion signals are organized into semantic descriptions, providing interpretable evidence across modalities.
\textbf{(2) Emotion cue fusion and personality assessment module.} In this module, the extracted emotion cues are integrated with interview responses and incorporated into MLLMs-based personality assessment through structured reasoning. This design enables the model to jointly reason over verbal content and nonverbal behavioral cues for fine-grained personality assessment.
The overall framework is illustrated in Fig.~\ref{fig.full_framework}. Details of each component are described below.

\subsection{Preprocessing}
Our method focuses on leveraging multimodal emotion signals as the core pathway for personality assessment. Thus we decompose the interview videos into independent visual and audio modalities. The decomposition allows the model to separately capture facial expressions, sound, and linguistic content from different modalities. The audio stream is further transcribed into text to ensure semantic alignment across modalities. This independent processing of different modalities enables MLLMs to jointly model both verbal and non-verbal cues. Thus, our method can lay the foundation for multimodal emotion representation and support subsequent personality assessment.

\subsection{Emotion cue extraction stage}
\begin{figure}[htbp]
\centering
\includegraphics[width=\linewidth]{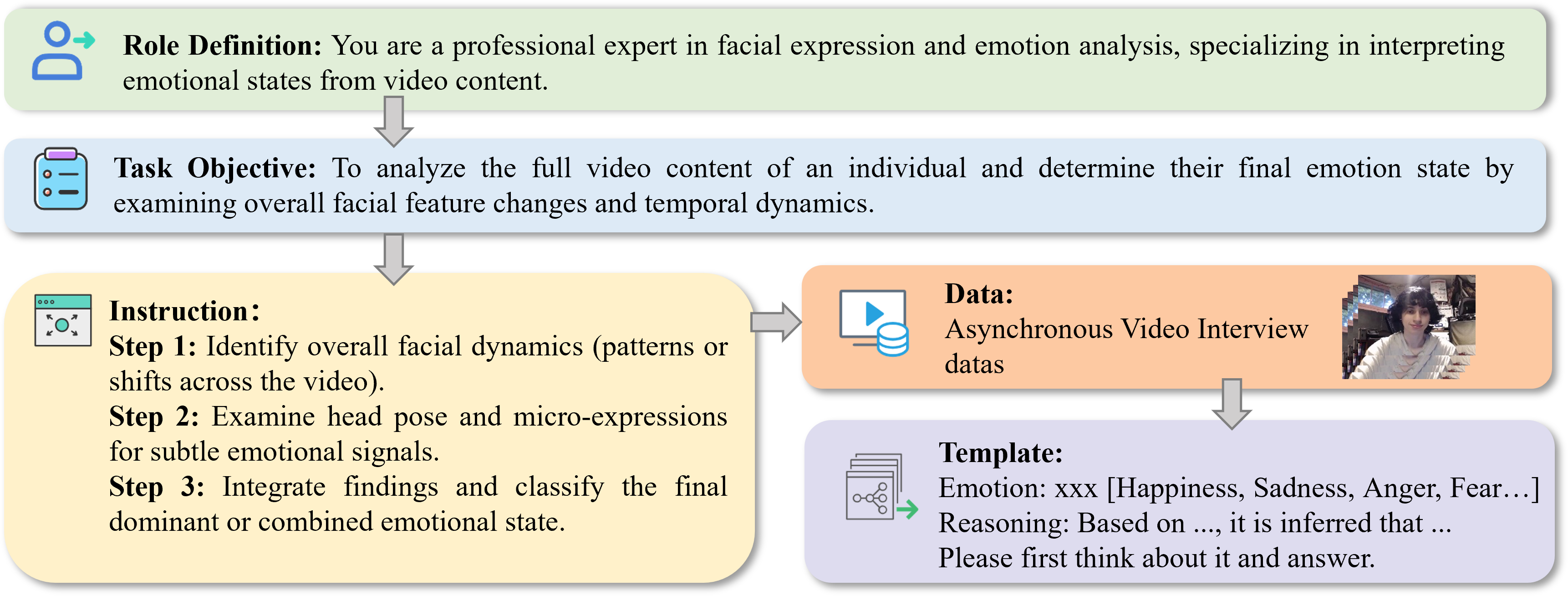} 
\caption{Prompt for emotion cue extraction stage} 
\label{fig:prompt_emotion}
\end{figure}

To enable MLLMs to extract comprehensive emotion cues from multimodal data, we design a structured prompting framework that emphasizes interpretability and stepwise analysis. The prompt consists of five components: 1) \textbf{Role Definition}, 2) \textbf{Task Objective}, 3) \textbf{Instructions}, 4) \textbf{Data Input} and 5) \textbf{ Answer Template}. The overall structure is illustrated in Fig.~\ref{fig:prompt_emotion}.

\begin{figure*}[htbp]
\centering
\includegraphics[width=1\textwidth]{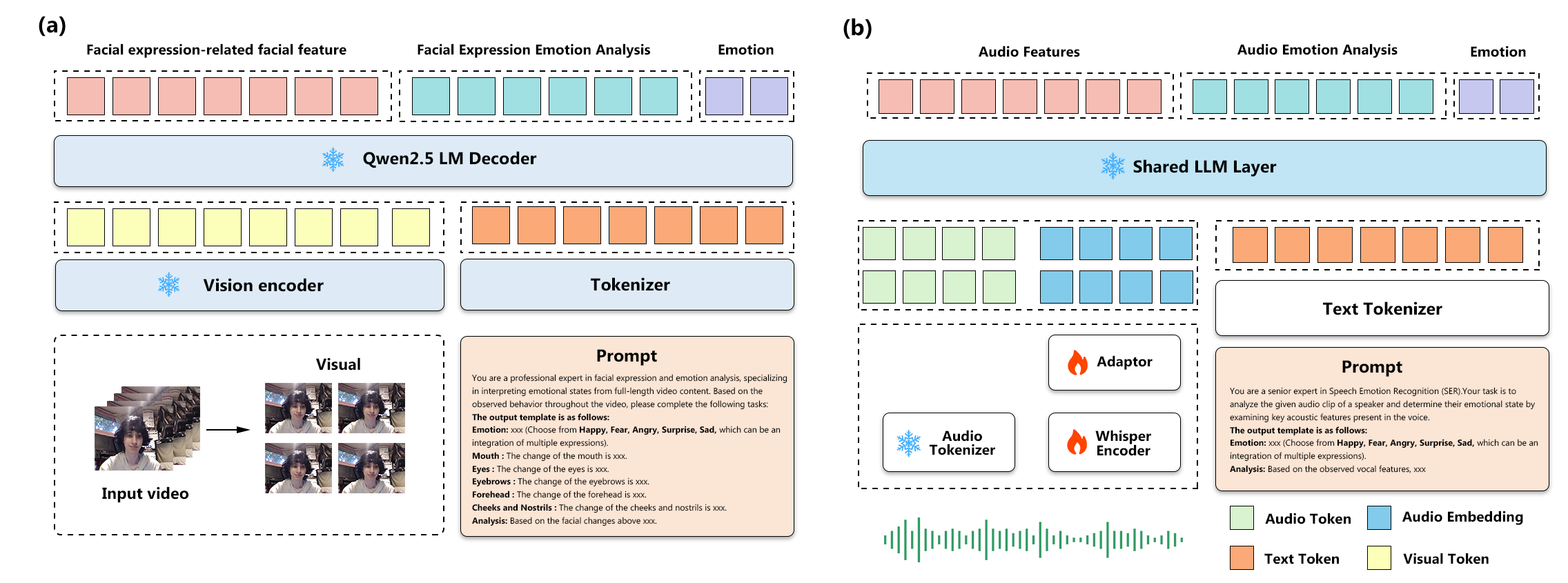}
\caption{Implementation of the emotion cue extraction stage. (a) Emotion cue extraction from the visual modality; (b) Emotion cue extraction from the audio modality}
\label{fig:vllm+audio_frame}
\end{figure*}

\begin{enumerate}
  
\item \textbf{Role Definition:} The model is assigned the role of a multimodal emotion analysis expert, responsible for analyzing facial expressions, vocal characteristics, and textual sentiment. This role setting encourages the model to perform analysis from a unified perspective that combines behavioral and linguistic cues.

\item \textbf{Task Objective:} The primary goal of this prompt is explicitly defined as the extraction of emotion cues from visual and audio modalities, followed by an analysis of their temporal dynamics. The final output includes emotion cues such as anger, happiness, or neutrality.

\item \textbf{Instructions:} This part uses an explicit chain-of-thought strategy\cite{10537616} to guide the model to execute emotion reasoning process.
\begin{itemize}

\item \textbf{Step 1:} “\textit{First, identify the modality type (visual or audio modality) and extract corresponding timestamp information}”: this step establishes a modality-aware framework in which different data streams are routed to their dedicated processing pathways, with alignment mechanisms ensuring proper temporal synchronization.

\item \textbf{Step 2:} “\textit{Then analyze the dynamic changes within each modality, for visual, observe facial action units in regions like eyebrows, eyes, and mouth; for auditory, examine prosodic features such as pitch variation and speech rhythm}”:this step enables fine-grained extraction of modality-specific behavioral and emotion cues.

\item \textbf{Step 3:} “\textit{Finally, generate emotion state and provide a traceable explanation of how the observed cues, such as ‘raised eyebrows’ or ‘rising intonation’ led to the inferred emotion}”– this step integrates the preliminary analysis results. The model outputs the emotion state. It also gives a traceable reasoning process based on the observed cues to enhance interpretability.
\end{itemize}
This stepwise design encourages the model to ground emotion inference in observable multimodal behaviors, improving the consistency and interpretability of extracted emotion cues.

\item \textbf{Data Input:} The input consists of a series of question–answer videos or audio recordings. Each question is associated with a specific HEXACO personality trait. The question design aims to elicit responses that contain corresponding emotion cues.

\item \textbf{Answer Template:} The prompt instructs the model to follow the answer template to respond. It also encourages the model to produce responses after sufficient analysis, improving the reliability of the extracted emotion cues.

\end{enumerate}

After the initial extraction, the emotion cues are further refined to improve their consistency and reliability. We use an LLM (Gemma2\cite{team2024gemma}) to perform structured filtering on the raw emotion cues. Specifically, the model is prompted to (1) identify duplicated or semantically overlapping cues (e.g., 'smiling lightly' and 'slightly raised mouth corners') and retain only the most representative expression; and (2) remove vague or uncertain judgments (e.g., 'might be nervous' and 'seems somewhat happy') by requiring explicit behavioral evidence. During this refinement process, Gemma2 preserves modality tags for each emotion cue while filtering out redundant and ambiguous descriptions. The resulting set of emotion cues is more consistent and interpretable, providing reliable input for the subsequent personality assessment stage.

\subsection{Emotion cue integration and personality assessment stage}

\begin{figure}[htbp]
\centering
\includegraphics[width=\linewidth]{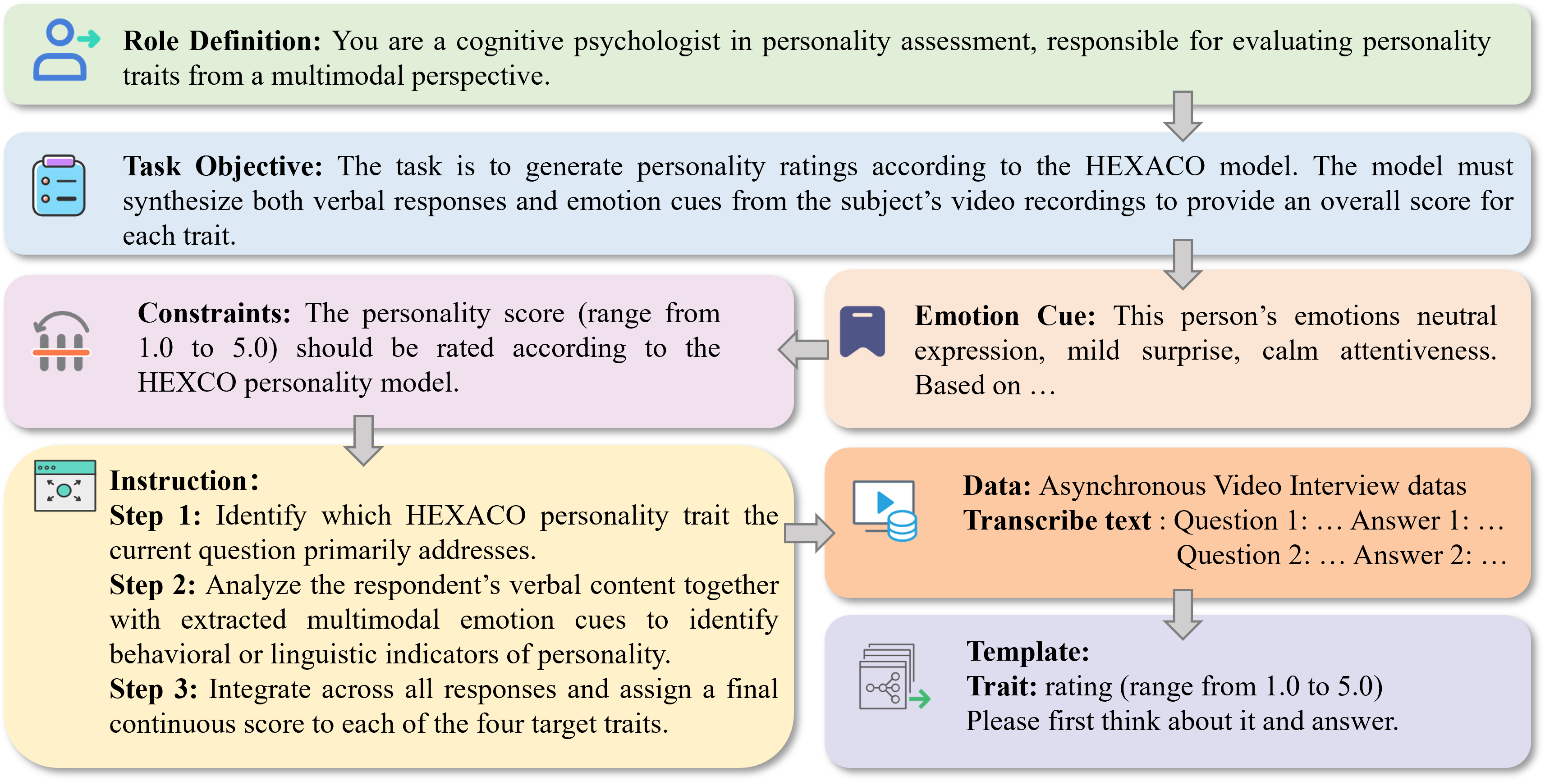} 
\caption{Prompt for personality assessment stage} 
\label{fig:prompt_personality}
\end{figure}

To support trait-level personality assessment, we design a structured prompting framework in the second stage of \textit{EMMR}. Building on the extracted emotion cues, this framework integrates emotion-related behavioral evidence with interview responses and incorporates them into MLLMs-based reasoning.
Similar to the emotion cue extraction module, the prompt is composed of seven core components: 1) \textbf{Role Definition}, 2) \textbf{Task Objective}, 3) \textbf{Emotion Cue Label} (The functions and influences of different emotion cue labels will be elaborated in detail in discussion.), 4) \textbf{Constraints}, 5) \textbf{Instructions}, 6) \textbf{Data Input}, and 7)\textbf{ Answer Template}. As illustrated in Fig.\ref{fig:prompt_personality}, the extracted emotion cues are explicitly included as auxiliary behavioral evidence alongside interview responses, enabling the model to process multimodal information within a unified reasoning context.

The prompt design in this stage follows a structure similar to that used in the emotion cue extraction stage but remains distinct in objective and functional emphasis. While the previous stage primarily focused on identifying affective states from multimodal inputs, the present stage shifts toward personality assessment. In this stage, the model is assigned the role of a cognitive psychologist in personality assessment, leading MLLMs to approach the task from the theoretical perspective of personality psychology. The goal of task for this stage is to transform the observed emotion signals into psychologically meaningful evidence that supports fine-grained personality assessment. The fine-grained prompt requires the model to assign a continuous numeric score (ranging from 1.0 to 5.0) for each trait. 

Another distinction of the fine-grained prompt lies in the inclusion of prior information in the form of emotion cue labels. Each input not only contains the participant’s video and question–answer pairs but also incorporates a pre-assigned emotion cues (e.g., calm, nervous) along with the underlying cause of that emotion, inferred from the multimodal analysis conducted in the previous stage. These cues include both the type of emotion and its observable triggers in behavior. For example, enthusiastic indicated by smiling and widened eyes. Such emotion cues serve as auxiliary signals that help the model anchor its reasoning in multimodal behavioral evidence rather than relying solely on text-based semantics. By integrating these emotion-based priors, the model can capture subtle affective nuances and produce more psychologically consistent and interpretable personality assessment.

The structured instructions guide the model to perform stepwise reasoning for personality assessment. Specifically, the model first identifies the target personality trait associated with each question, then analyzes the participant’s responses together with multimodal emotion cues to extract relevant behavioral evidence, and finally assigns a continuous score with justification grounded in the observed cues. This process establishes a connection between observable emotion-related behaviors and HEXACO personality traits.

By integrating emotion cues extracted from multimodal data as intermediate bridges between behavioral signals and personality traits, personality assessment can be performed not only based on explicit textual content but also on implicit non-verbal emotion expressions. Thereby overcoming the limitations of single-modal assessment that relies solely on text information. Based on the above two stages, the framework can effectively achieve fine-grained, interpretable, and consistent personality assessment. Experiments show that the proposed method is superior to the traditional LLMs-based single-modal personality assessment, verifying its effectiveness.

\section{Datasets}

To evaluate the effectiveness of \textit{EMMR}, we conduct experiments on two AVI-based personality assessment datasets, OPVA \cite{koutsoumpis2024beyond} and AVI-6 \cite{dataset2025}. Although both datasets are developed under the HEXACO personality framework, they provide annotations for only a subset of HEXACO dimensions. Therefore, our experiments are conducted on the personality traits available in each dataset. The following is a detailed description of the two datasets.

The OPVA dataset \cite{koutsoumpis2024beyond} contains AVIs recordings from 685 candidates involved in a simulated management traineeship application process. To focus on workplace-relevant personality traits, psychologists designed eight interview questions targeting two HEXACO dimensions: eXtraversion (X) and Conscientiousness (C), with four questions associated with each trait. These two traits were selected because prior studies have shown that they are strong predictors of workplace performance\cite{tett2021trait}.
The AVI-6 dataset \cite{dataset2025} includes AVIs recordings from 646 candidates engaged in simulated job application interviews. Its interview questions designed cover multiple HEXACO personality dimensions. Psychologists developed four questions to elicit expressions of Honesty-Humility (H), eXtraversion (X), Agreeableness (A), and Conscientiousness (C). In addition, the dataset contains two general interview questions commonly used in recruitment settings.

We selected these two datasets for several reasons. First, the interview questions in both datasets were explicitly designed by psychologists based on Trait Activation Theory \cite{tett2021trait}, ensuring that the target personality traits were intentionally elicited rather than implicitly inferred from unconstrained behaviors. In contrast, many publicly available personality datasets \cite{ponce2016chalearn,sun2024revealing}, especially those collected from internet videos such as ChaLearn \cite{ponce2016chalearn} and PersonalityEvd \cite{sun2024revealing}, do not clearly specify which personality traits are activated in their recordings. Second, both datasets are constructed in simulated recruitment settings, making them more consistent with real AVI-based personality assessment scenarios and reducing potential biases introduced by non-interview contexts. Therefore, these datasets provide a more suitable benchmark for evaluating MLLMs-based personality assessment in AVIs.

\section{Experiments and results}

In this section, we evaluate \textit{EMMR} on two AVIs datasets and analyze its effectiveness for MLLMs-based personality assessment. We present the implementation details, evaluation metrics, main results, baseline comparisons, and additional analyses of emotion cues.

\subsection{Implementation details}

For each AVIs response, we first separate the original video into visual and audio streams using MoviePy \cite{10932863}. The audio stream is then transcribed into text using Google's Speech-to-Text technology. This splitting approach enables MLLMs to capture not only the semantic information embedded in text but also the non-verbal cues conveyed through audio, thereby laying the foundation for multimodal emotion modeling and supporting personality assessment. We used the open-sourced medium-sized model (around 10B parameters) Qwen2.5-VL(7B)\cite{bai2025qwen2} as the backbone MLLMs for \textit{EMMR}. Due to their limited number of parameters, small-sized models (around 2B parameters) often show weak generalization according to the scaling laws. Deploying large-sized models (with over 40B parameters) requires high hardware specifications. To balance computational efficiency and prediction performance, we selected the medium-sized model. 

%We chose the open-sourced MLLMs because we could get rid of the access and usage cost limitations of closed-sourced models. In addition, the open-sourced model allows modification of the model structure and inference process to better adapt to the task characteristics of multimodal emotion cue extraction and personality assessment.

We implement the \textit{EMMR} framework using a modular multi-model architecture. Specifically, Qwen2.5-VL (7B) \cite{bai2025qwen2} is employed for visual emotion cue extraction and downstream personality trait assessment due to its strengths in vision-language understanding. For audio emotion analysis, Kimi-Audio (7B)\cite{ding2025kimi} is utilized to process the intonation feature information form audio inputs. We used Gemma2 (9B)\cite{team2024gemma} as the cue integration module to refine emotion signals from each modality and convert them into a clean, unified text format. All experiments and implementations are carried out on a desktop computer equipped with vGPU (48GB).

\subsection{Evaluation metrics}
For the assessment of personality trait predictions, we used three widely used metrics: Mean Absolute Error (MAE) , Mean Squared Error (MSE) and Pearson Correlation Coefficient (PCC). MAE measures the average magnitude of absolute differences between predicted and true trait scores, without considering direction. It is defined as:

\begin{equation}
MAE=\frac{1}{n}\sum_{i=1}^{n}\left|y_i-x_i\right|,\label{eq1}
\end{equation}
where $n$ represents the total number of samples, $y_i$ represents the true personality rating of the $i$th sample, and $x_i$ represents the rating predicted by the model. MSE quantifies the average of the squared differences between predicted and true values. It is defined as:
\begin{equation}
MSE=\frac{1}{n}\sum_{i=1}^{n}\left(y_i-x_i\right)^2,\label{eq2}
\end{equation}
using the same notation as above. MSE emphasizes larger errors due to the squaring operation, making it more sensitive to significant deviations and thus useful for detecting models that produce high-variance predictions.
PCC evaluates the linear relationship between predicted scores and human-annotated ground truth. It is defined as:
\begin{equation}
PCC=\frac{\sum_{i=1}^{n}\left(x_i-\bar{x}\right)\left(y_i-\bar{y}\right)}{\sqrt{\sum_{i=1}^{n}\left(x_i-\bar{x}\right)^2}\sqrt{\sum_{i=1}^{n}\left(y_i-\bar{y}\right)^2}}.\label{eq}
\end{equation}

$\bar{x}$ and $\bar{y}$ are the averages of the predicted and true values, respectively. $PCC\in\left[-1,1\right]$, the closer the PCC value is to 1, the stronger the linear correlation between the predicted results of the regression task optimization model and the true label, and the predicted trend is more consistent with the actual situation.

All metrics are extensively used in personality assessment tasks where continuous trait scores are predicted. In our experiments, the model outputs fine-grained, continuous trait ratings derived from multimodal emotion cues. These evaluation metrics are suited for evaluating the accuracy of personality assessment. 
%MAE clearly reflects the average level of prediction bias, MSE helps to identify outlier predictions that are likely to deviate significantly from human judgment, while PCC measures the correlation of model predictions with the true labels.

\begin{table*}[ht]
\centering
\scriptsize
\caption{The results of \textit{EMMR} and baseline methods}
\label{tab:model_perf}
\setlength{\tabcolsep}{6pt}
\renewcommand{\arraystretch}{1.25}

\resizebox{\textwidth}{!}{
\begin{tabular}{|c|c|c|c|c|c|c|c|c|}
\hline
\multicolumn{9}{|c|}{\textbf{Mean Absolute Error (MAE)}} \\ \hline
\multirow{2}{*}{\textbf{Models}} &
\multicolumn{3}{c|}{OPVA} &
\multicolumn{5}{c|}{AVI-6} \\ \cline{2-9}
 & X & C & Avg & H & A & X & C & Avg \\ \hline

Qwen2.5\cite{hui2024qwen2} & 0.4495 & 0.4142 & 0.4318 & 0.3787 & 0.3819 & 0.6308 & 0.4184 & 0.4524 \\ 
\hline
Kimi-audio\cite{ding2025kimi} & 0.5638 & 0.3244 & 0.4441 & 0.3287 & 0.4487 & 0.5707 & 0.4136 & 0.4404 \\
\hline
Kimi-VL\cite{team2025kimi} & 0.3968 & 0.3427 & 0.3697 & 0.4809 & 0.4043 & 0.5246 & 0.4381 & 0.4620 \\
\hline
Qwen2-VL\cite{wang2024qwen2} & 0.8019 & 0.4261 & 0.6140 & 0.3796 & 0.3118 & 0.5066 & 0.4158 & 0.4034 \\
\hline
Qwen2.5-VL\cite{bai2025qwen2} & 0.5989 & 0.8074 & 0.7030 & \textbf{0.2317} & 0.3456 & 0.3213 & 0.3698 & 0.3171 \\
\hline
DAN\cite{wei2017deep} & 0.6905 & 0.5187 & 0.6046 & 0.5172 & 0.4299 & 0.3755 & 0.3470 & 0.4156 \\
\hline
VAT\cite{girdhar2019video} & 0.6571 & 1.0486 & 0.8528 & 0.4277 & 0.5103 & 0.5054 & 0.5397 & 0.4957 \\
\hline
Swin-transformer\cite{liu2021swin} & 0.6131 & 0.4607 & 0.5369 & 0.2926 & 0.3185 & 0.3247 & 0.2827 & 0.3046 \\
\hline
\textbf{EMMR} & \textbf{0.3686} & \textbf{0.2872} & \textbf{0.3279} & 0.3373 & \textbf{0.3138} & \textbf{0.3114} & \textbf{0.2553} & \textbf{0.3045} \\
\hline

\multicolumn{9}{|c|}{\textbf{Mean Squared Error (MSE)}} \\ \hline
Qwen2.5\cite{hui2024qwen2} & 0.3200 & 0.2533 & 0.2866 & 0.2351 & 0.2230 & 0.5493 & 0.2640 & 0.3179 \\
\hline
Kimi-audio\cite{ding2025kimi} & 0.4763 & 0.1599 & 0.3181 & 0.1623 & 0.3432 & 0.4495 & 0.2572 & 0.3030 \\
\hline
Kimi-VL\cite{team2025kimi} & 0.2445 & 0.1804 & 0.2125 & 0.3527 & 0.2440 & 0.4426 & 0.2873 & 0.3317 \\
\hline
Qwen2-VL\cite{wang2024qwen2} & 0.9514 & 0.2801 & 0.6157 & 0.1942 & 0.1580 & 0.3321 & 0.2445 & 0.2322 \\
\hline
Qwen2.5-VL\cite{bai2025qwen2} & 0.5437 & 0.8542 & 0.6990 & \textbf{0.0822} & 0.1676 & 0.2015 & 0.1968 & 0.1620 \\
\hline
DAN\cite{wei2017deep} & 0.7240 & 0.4041 & 0.5641 & 0.3837 & 0.2425 & 0.1985 & 0.1813 & 0.2515 \\
\hline
VAT\cite{girdhar2019video} & 0.6736 & 1.3507 & 1.0121 & 0.2227 & 0.3315 & 0.3087 & 0.3752 & 0.3095 \\
\hline
Swin-transformer\cite{liu2021swin} & 0.5577 & 0.3409 & 0.4493 & 0.1642 & 0.1822 & 0.1667 & 0.1319 & 0.1612 \\
\hline
\textbf{EMMR} & \textbf{0.2109} & \textbf{0.1310} & \textbf{0.1710} & 0.1606 & \textbf{0.1501} & \textbf{0.1533} & \textbf{0.0928} & \textbf{0.1392} \\
\hline

\multicolumn{9}{|c|}{\textbf{Pearson Correlation Coefficient (PCC)}} \\ \hline
Qwen2.5\cite{hui2024qwen2} & 0.7185 & 0.6002 & 0.6594 & 0.1563 & 0.4745 & -0.0871 & 0.5683 & 0.2780 \\
\hline
Kimi-audio\cite{ding2025kimi} & 0.7735 & 0.7082 & 0.7408 & 0.2730 & 0.4625 & -0.0888 & \textbf{0.6027} & 0.3123 \\
\hline
Kimi-VL\cite{team2025kimi} & 0.7521 & 0.6195 & 0.6858 & 0.1709 & 0.4453 & -0.1606 & 0.5227 & 0.2446 \\
\hline
Qwen2-VL\cite{wang2024qwen2} & 0.0850 & 0.0946 & 0.0898 & 0.0244 & 0.1247 & -0.0486 & 0.1684 & 0.0672 \\
\hline
Qwen2.5-VL\cite{bai2025qwen2} & 0.2230 & 0.1117 & 0.1674 & 0.1081 & 0.1481 & -0.0282 & 0.2302 & 0.1145 \\
\hline
DAN\cite{wei2017deep} & 0.0472 & 0.0572 & 0.0522 & -0.0795 & -0.0367 & -0.0995 & 0.0748 & -0.0352 \\
\hline
VAT\cite{girdhar2019video} & -0.1359 & -0.0143 & -0.0751 & 0.0413 & -0.0156 & 0.1004 & 0.0950 & 0.0553 \\
\hline
Swin-transformer\cite{liu2021swin} & 0.0659 & 0.0269 & 0.0464 & -0.0392 & -0.0677 & 0.0501 & 0.0503 & 0.0190 \\
\hline
\textbf{EMMR} & \textbf{0.8338} & \textbf{0.7432} & \textbf{0.7885} & \textbf{0.2864} & \textbf{0.5193} & \textbf{0.3527} & 0.6004 & \textbf{0.4397} \\
\hline

\end{tabular}}
\end{table*}     

\subsection{Results}
As shown in Table~\ref{tab:model_perf}, \textit{EMMR} achieves an MAE of 0.3686 for eXtraversion (X) and 0.2872 for Conscientiousness (C), alongside MSE of 0.2109 and 0.1310 respectively, on the OPVA dataset. Because OPVA provides dense information per trait (i.e., four corresponding questions each), \textit{EMMR} can leverage the explicit emotion semantics to corroborate and enrich the verbal behavioral profile. The resulting low MAE and MSE values suggest that introducing these structured emotion descriptions effectively helps prevent large assessment errors, particularly for structured traits like Conscientiousness.

In contrast, the AVI-6 dataset covers a broader range of personality dimensions but provides only one question per trait, resulting in sparsely distributed information. Under these low-context conditions, \textit{EMMR} yields an MSE of 0.1606 for Honesty-Humility (H), 0.1501 for Agreeableness (A), 0.1533 for X, and 0.0928 for C, culminating in a low average MSE of 0.1392. These observable error reductions across both datasets indicate that explicitly incorporating multimodal emotion cues into the reasoning process effectively suppresses prediction noise, thereby enhancing the overall reliability of the assessment even when verbal context is limited.

We also evaluate the PCC. \textit{EMMR} achieves a high PCC of 0.8338 for X and 0.7432 for C on OPVA, demonstrating that using emotion cues semantics as behavioral anchors helps the method maintain strong rank consistency with human judgments when sufficient trait-specific context is available. On AVI-6, although a reasonable level of consistency is maintained, the correlation is noticeably lower than on OPVA. This decline can be attributed to the reduced number of task-related inputs, which makes it harder to capture stable behavioral patterns. This weaker correlation underscores that while emotion cues semantics provide valuable supplementary evidence, baseline response richness remains fundamental for accurate personality modeling; traits become inherently harder to identify when fewer cues are available.

Overall, these results illustrate that the emotion-mediated reasoning approach in \textit{EMMR} adapts well to both dense (OPVA) and sparse (AVI-6) data environments. The assessment performs optimally when abundant verbal background information can be cross-referenced with emotion cues semantics. Yet, even in lower-context conditions, \textit{EMMR} can still maintain competitive accuracy and robustness.

\subsection{Comparison with baselines}
To evaluate the effectiveness of \textit{EMMR}, we compare it against two distinct categories of baseline models. The first category comprises LLMs at the medium-sized (7B) scale, ensuring fair comparability of computational resources. This selection includes a text-only baseline (Qwen2.5\cite{hui2024qwen2}), an audio-centric model (Kimi-audio\cite{ding2025kimi}), and three visual-language models (Kimi-VL\cite{team2025kimi}, Qwen2-VL\cite{wang2024qwen2}, and Qwen2.5-VL\cite{bai2025qwen2}). As shown in Table~\ref{tab:model_perf}, these modern MLLMs demonstrate substantial instability across different traits when not explicitly guided by structured emotion cues. For instance, Qwen2.5-VL\cite{bai2025qwen2} achieves a low MAE of 0.2317 on Honesty-Humility (H), but its MAE surges to 0.8074 on C within the OPVA dataset. A similar variability is observed for Qwen2-VL\cite{wang2024qwen2}, which performs well on A (MAE = 0.3118) yet experiences a sharp error increase on X (MAE = 0.8019).

We also evaluate three representative visual models from the pre-MLLMs era: DAN \cite{wei2017deep}, VAT \cite{girdhar2019video} and Swin-transformer \cite{liu2021swin}. These classical models, which emphasize spatial and temporal feature learning, were trained on a subset of the dataset (120 samples\cite{11481797}, approximately 20\%) and tested on the remaining 80\%. While these models achieve moderate MAE and MSE in certain cases, such as Swin-Transformer attaining a relatively strong average MSE of 0.1612 on AVI-6, they exhibit remarkably weak PCC. Their PCC values are confined to a narrow range between –0.1359 and 0.0950. This discrepancy highlights that although traditional neural networks can minimize numerical loss on specific datasets, their black-box feature extraction aligns poorly with human judgments, failing to capture the high-level semantic nuances necessary for reliable personality assessment.

These results demonstrate that\textit{ EMMR} consistently outperforms all baseline models across evaluation metrics and datasets. While some baselines have shown advantages in particular traits, their generalizability and stability is weak (i.e., unstable performance among different traits). In contrast, our method can conduct robust and human-aligned personality assessment under both rich and sparse input conditions due to multimodal emotion cues semantics fusion. Overall, \textit{EMMR} proves to be an effective approach forMLLMs-based personality assessment in AVIs.

\subsection{Ablation study}
\begin{figure}[htbp]
\centering
\includegraphics[width=\linewidth]{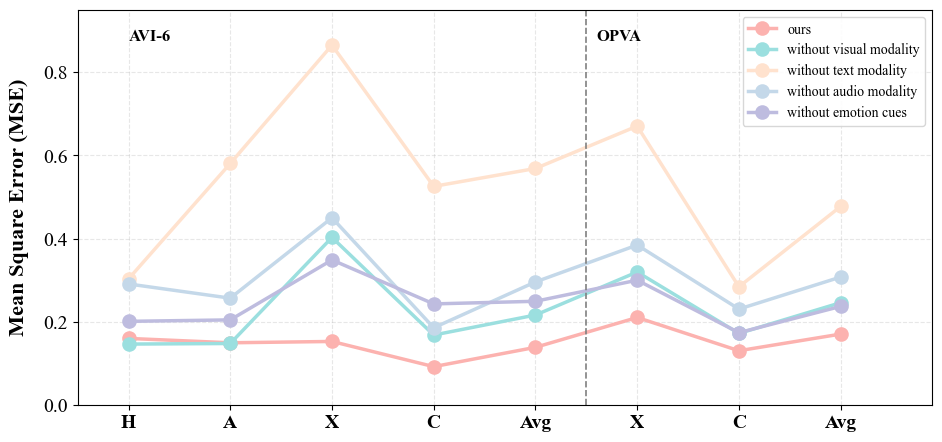}
\caption{The results of ablation study} 
\label{fig:ablation}
\end{figure}

To verify the effectiveness of our proposed method, we conduct a comprehensive ablation study by removing key input components from the framework. The experimental results are shown in Fig.~\ref{fig:ablation}. The complete framework integrates three modalities, video, transcribed text of question-answer pairs, and multimodal emotion cues. The evaluated variants include:
(1) \textbf{without visual modality}, excluding video input; (2) \textbf{without text modality}, excluding the transcribed responses; (3) \textbf{without audio modality}, excluding acoustic information; and (4) \textbf{without emotion cues}, removing the extracted semantic representations of emotion. The purpose of this is to help clarify the contribution of each component to the overall personality assessment performance.

The results demonstrate that visual, audio, and emotion cues each play essential and complementary roles in multimodal personality assessment. When the visual modality is removed, the model’s average MSE increases from 0.1710 to 0.2464 on OPVA and from 0.1503 to 0.2168 on AVI-6, indicating a clear performance drop for personality assessment. Removing the audio modality also leads to a noticeable increase in MSE (0.3080 on OPVA and 0.2963 on AVI-6). Similarly, excluding emotion cues results in an average MSE increase of 0.2498 on OPVA and 0.2375 on AVI-6. The performance decrease suggests that emotion-derived information effectively bridges behavioral signals and personality traits.

The removal of transcribed text causes a substantial degradation in model performance, underscoring its foundational role in MLLMs-based personality assessment. Without transcribed text, the average MSE surges to 0.4776 on OPVA and 0.5688 on AVI-6, which demonstrates the increases of over 178\% and 278\% respectively compared to our method.  In particular, the MSE for Agreeableness rises to 0.5808 and for Extraversion to 0.8650 on AVI-6, indicating a severe loss in predictive capability for these traits. This decline implies that transcribed text is not merely auxiliary; it provides explicit semantic meaning, contextual grounding, and logical structure that are highly relevant for interpreting personality-revealing behaviors. While visual and emotion cues enrich the model’s understanding, they do not appear to fully compensate for the absence of linguistic semantics.
In summary, the ablation experiments suggest that each modality contributes positively to the overall performance, and their integration yields the most stable assessment results within our experiments.

\section{Discussion}

\subsection{Which modal information is most suitable for assisting personality assessment?}
From a psychological perspective, different modalities play distinct roles in personality expression. Textual information directly reflects an individual’s cognitive and language trait manifestations, while visual and audio modalities convey non-verbal emotion and behavior cues.  However, it remains an open question which modality serves as the most effective auxiliary signal for automated assessment. To address this gap, we proposed research question 1: Which modal information is most suitable for assisting personality assessment?

\begin{figure}[h!]
\centering
\includegraphics[width=\linewidth]{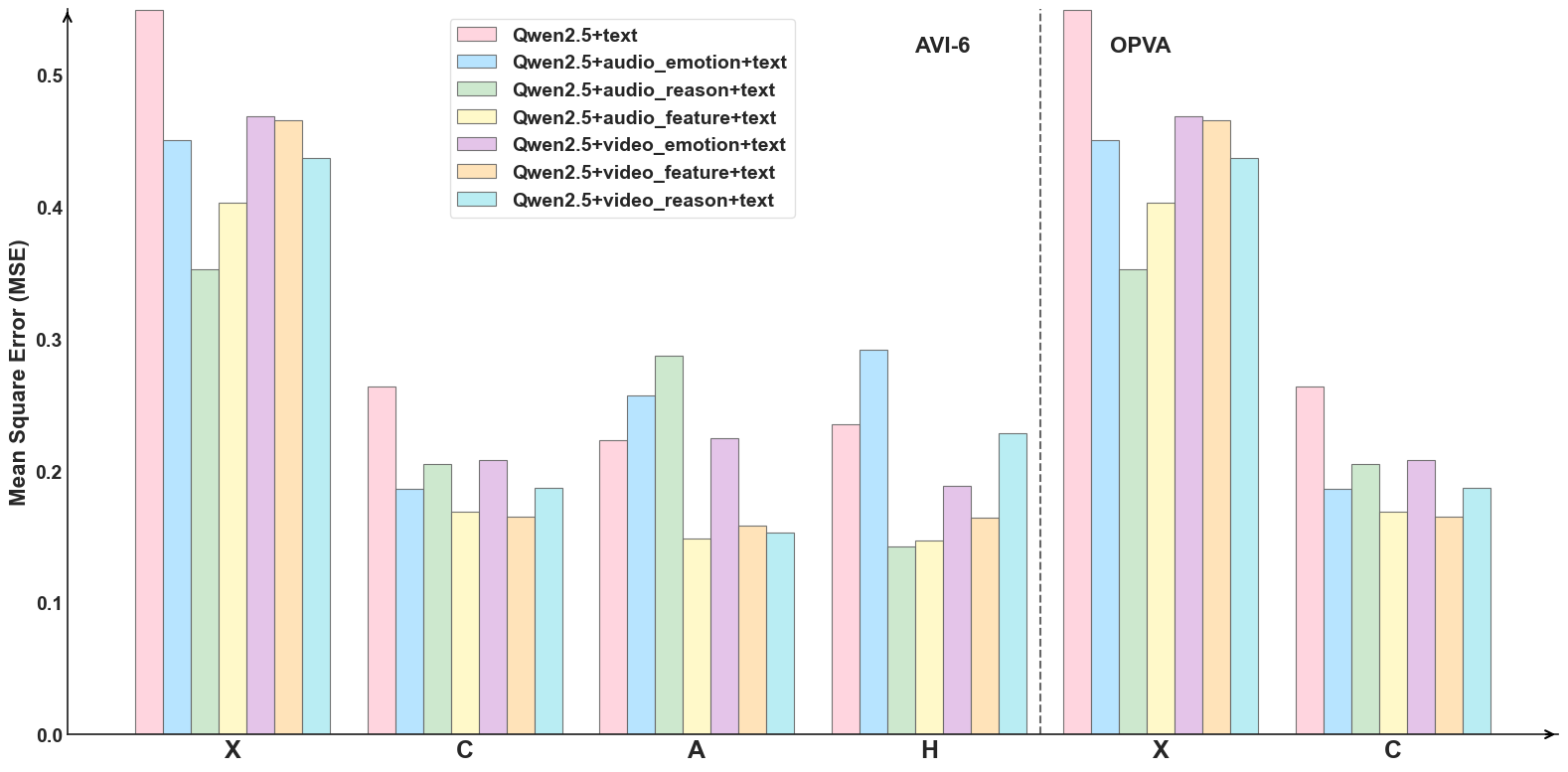} 
\caption{The comparison between different backbone model} 
\label{fig:model}
\end{figure}

To answer this question, we designed two sets of comparative experiments based on MSE. The first set evaluated the independent effectiveness of single-modal inputs (text-only, visual-only, audio-only). As shown in Fig.~\ref{fig:model}, the results reveal distinct characteristics for each modality. The text-only baseline (Qwen2.5\cite{hui2024qwen2}) demonstrates stable performance across datasets, with average MSE values of 0.2866 on OPVA and 0.3179 on AVI-6. In contrast, the visual-only model (Qwen2.5-VL\cite{bai2025qwen2}) exhibits extreme instability, performing poorly on OPVA (MSE = 0.6990) but well on AVI-6 (MSE = 0.1620). The audio-only method (Kimi-audio\cite{ding2025kimi}) achieves a moderate average MSE of 0.3181 on OPVA, but shows noticeable disadvantages on AVI-6 (MSE = 0.3030).

The second set of experiments further evaluated the contribution of these modalities when structured semantic descriptions of their corresponding emotion cues were incorporated into the personality assessment. Compared to the text-only baseline, integrating explicit semantic descriptions of emotion improves both accuracy and stability, reducing the average MSE to 0.2464 on OPVA and 0.2168 on AVI-6. Conversely, incorporating semantic descriptions of audio emotion cues introduces instability; performance drops sharply on AVI-6, with the MSE increasing to 0.3677. This is likely because the inherent noise in acoustic emotion extraction translates into ambiguous or misleading textual semantics. Notably, when the Qwen2.5-VL model\cite{bai2025qwen2} is jointly prompted with transcribed text and semantic descriptions of visual emotion cues, it achieves the most stable and competitive results. Specifically, the integration of these visual emotion semantic descriptions yields the lowest MSE (0.1503) on AVI-6 and facilitates a substantial error reduction on OPVA (from 0.6990 down to 0.2394), underscoring the vital role of the semantic features of such emotion cues in enhancing the robustness of personality assessment.

These findings address our first research question by revealing that there is no absolute 'optimal' auxiliary modality in isolation; rather, their suitability depends heavily on the fusion strategy. While transcribed text serves as the indispensable semantic backbone, raw non-verbal signals introduce high variance when used independently. However, when volatile non-verbal behaviors are transformed into structured semantic descriptions of emotion cues and fused alongside the text, visual information becomes the most effective complementary modality. Ultimately, this semantic-level fusion mitigates the instability of single modalities by ensuring all behavioral evidence is aligned within the MLLM's reasoning space.

\subsection{Do semantic descriptions of emotion cues help?}

Psychological research\cite{9868797,marengo2021meta,sacchi2026understanding,9210819,zhu2023understanding,cai2025mdpe,jie2022impact,11458684} suggests that non-verbal emotion expressions exhibited during communication may convey rich behavioral signals relevant to stable individual differences. Features such as facial expressions, gestures, and vocal intonation are often associated with traits like Extraversion and Agreeableness. However, multimodal behavioral signals are inherently heterogeneous and difficult to directly incorporate into personality reasoning under the framework of MLLMs. Since MLLMs primarily operate in the semantic language space, these multimodal emotion-related behaviors typically need to be transformed into structured semantic descriptions to provide a more compatible representation for processing. Therefore, our second research question explores: to what extent do different semantic representations of multimodal emotion cues influence the model’s capability in personality assessment?

To investigate this, we construct three emotion variants from audio and visual modalities: (1) coarse-grained emotion label cues (e.g., happy, sad, neutral); (2) fine-grained audio or visual emotion label cues (e.g., pitch changes, facial features); (3) The model-generated emotion reasoning cues and their relationship with personality traits (e.g., moderate eye contact, occasional light smiles, and relatively neutral facial expressions, suggesting a balanced degree of social engagement and energy, indicative of a personality that is not strongly Extraversion ).

\begin{figure}[h!]
\centering
\includegraphics[width=\linewidth]{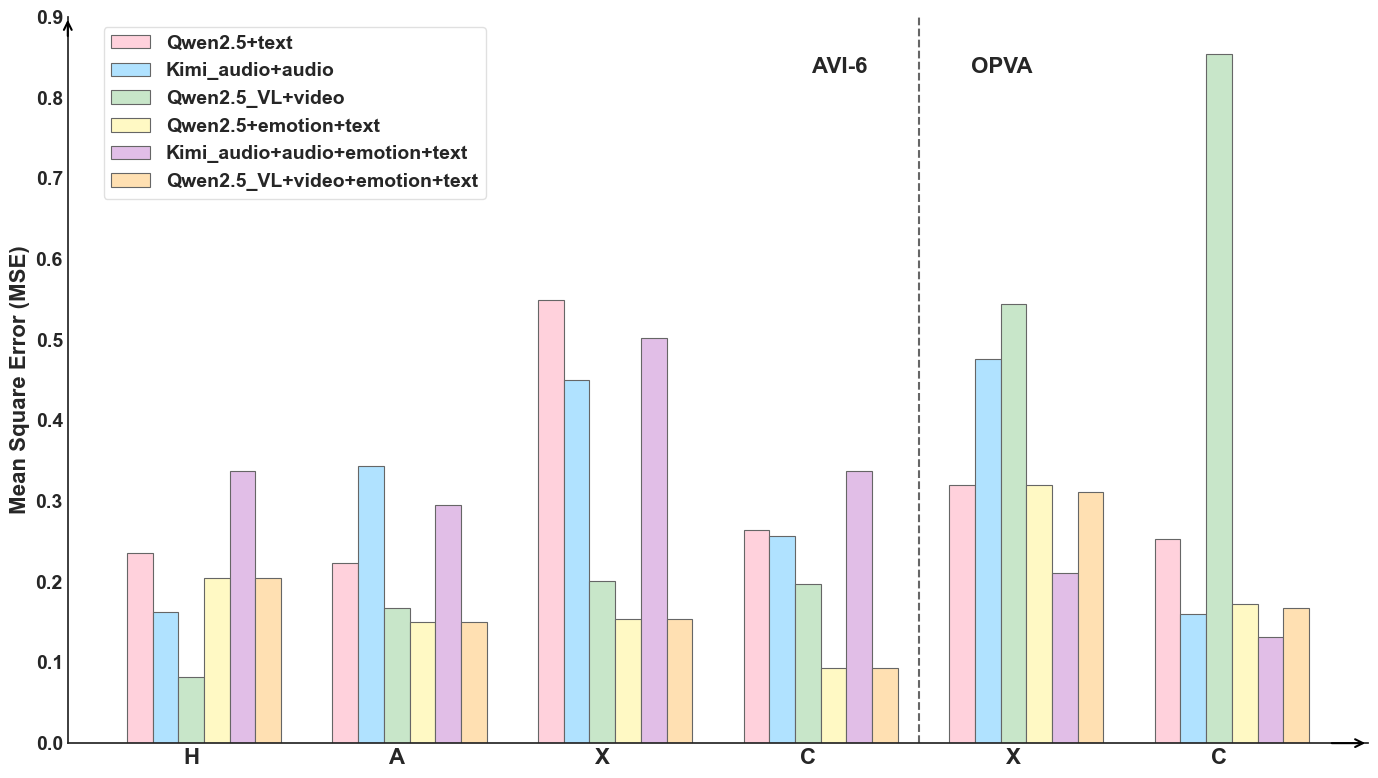} 
\caption{The comparison between different types of emotion cues} 
\label{fig:emotion}
\end{figure}

As shown in Fig.~\ref{fig:emotion}, integrating the semantics of emotion cues generally yields better performance compared to the text-only baseline(Qwen2.5\cite{hui2024qwen2}, average MSE 0.2866 on OPVA and 0.3179 on AVI-6). The inclusion of coarse-grained visual emotion semantics provides a moderate performance gain (average MSE 0.2530 and 0.2385). However, fine-grained semantic descriptions of visual features appear to introduce semantic noise, leading to observable performance drops, particularly on AVI-6 (average MSE rising to 0.3067). In contrast, the reasoning-based semantic descriptions of visual emotions demonstrate more stable improvements, reaching average MSEs of 0.2510 on OPVA and 0.2512 on AVI-6. Similarly, while coarse-grained audio emotion labels offer limited benefits (average MSE 0.3080 and 0.2963), the semantic reasoning of audio emotions further assists the assessment, bringing the average MSE down to 0.2464 on OPVA and 0.2467 on AVI-6.

These observations suggest that transforming non-verbal emotion information into appropriate semantic spaces can serve as a beneficial complement to pure dialogue transcripts. Among the different semantic representation strategies, emotion reasoning semantics seem to be the most compatible with the reasoning mechanisms of LLMs. This may be because reasoning semantics provide a high-level, logically coherent interpretation of behavioral signals, bridging the gap between raw multimodal expressions and psychological indicators. Coarse-grained labels offer limited descriptive richness, whereas fine-grained feature descriptions might introduce modality-specific fragmentation and semantic inconsistencies, potentially distracting the LLM from underlying personality patterns. Overall, the findings indicate that multimodal emotion information is more helpful when translated into interpretable, semantically rich reasoning rather than simple categorical labels or dis-jointed physical descriptions.

\begin{figure*}[t]
\centering
\includegraphics[width=1\textwidth]{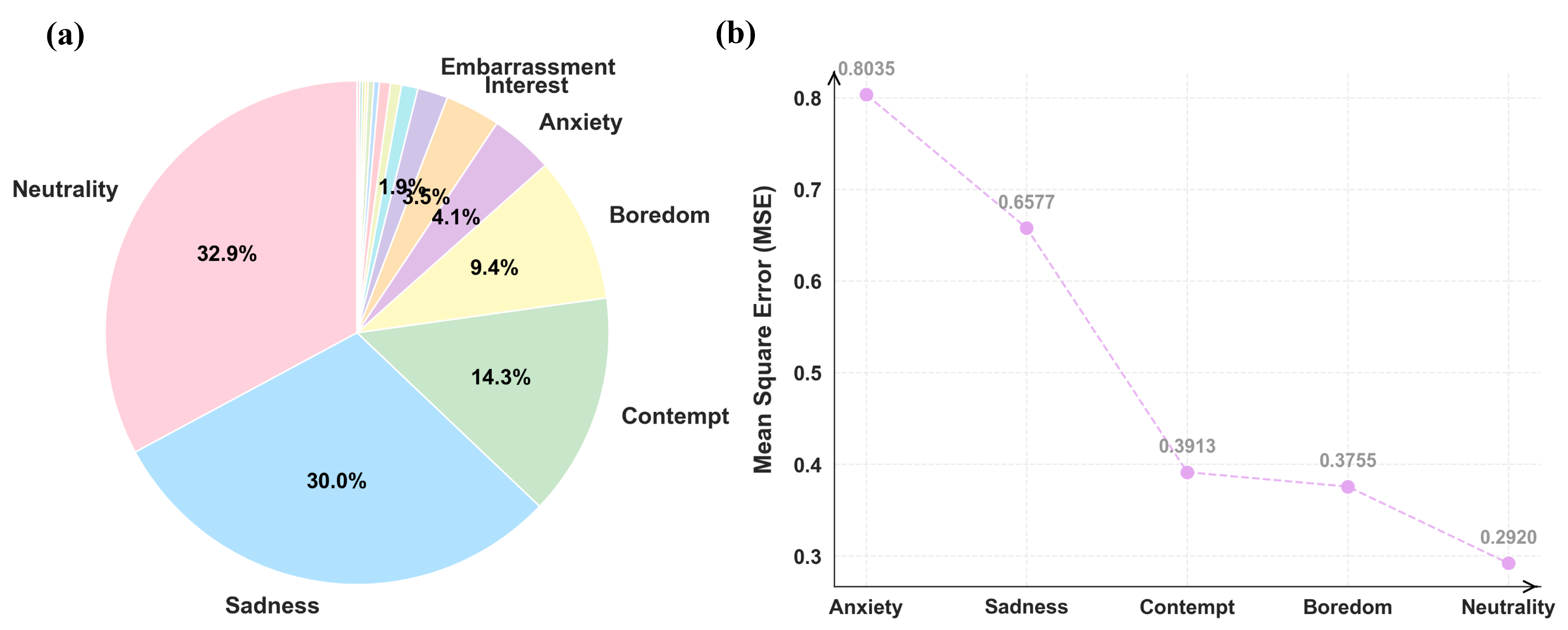}
\caption{The results of the Kimi-Audio model on the Extraversion. (a) The proportional distribution of emotion categories; (b) MSE corresponding to different emotion categories}
\label{fig:extraversion}
\end{figure*}

\subsection{Why does the inclusion of audio emotion cues increase the variability in Extraversion assessment?}

During our experiments, we observed that incorporating audio emotion cues introduced pronounced fluctuations in the assessment of the Extraversion trait. To investigate the underlying causes of this variability, we analyzed the distribution and impact of the extracted audio emotion states. As depicted in Fig.~\ref{fig:extraversion}(a), the emotion labels generated by the audio model indicate a substantial proportion of negative emotions, including sadness (30\%), contempt (14.3\%), boredom (9.4\%), and anxiety (4.1\%), neutral emotions account for 32.9\%. However, positive emotions(e.g., happiness) appear relatively scarce. When these emotion cues were integrated into the assessment model, the results (Fig.~\ref{fig:extraversion}(b)) revealed a distinct error pattern. Although neutral emotions were prevalent, they corresponded to the lowest prediction error (MSE=0.2920). In contrast, the assessment errors associated with negative emotion cues, particularly anxiety (MSE=0.8035) and sadness (MSE=0.6577), were notably larger.
 
Psychological literature\cite{duijndam2024behavior} suggests that negative emotions can inhibit outward individual behavior and social interaction, which are key manifestations of Extraversion. Our experimental observations align with this mechanism, albeit as an artifact of misclassification. Further inspection of the dataset revealed a critical discrepancy, the candidates’ actual emotional states were predominantly neutral. It appears that acoustic characteristics such as slower speech rates and lower pitches were frequently misclassified by the audio extraction models as negative emotions (e.g., sadness, contempt). When the MLLMs incorporates these erroneous negative semantic cues, it likely infers a lower level of social engagement, which directly degrades the accuracy of the Extraversion assessment and causes the observed fluctuations.

These misclassifications highlight the inherent complexity and ambiguity of mapping raw audio signals to discrete emotion categories\cite{10542112}. Overall, these findings underscore that the reliability of MLLMs-based personality assessment is highly dependent on the accuracy of upstream emotion cue extraction. Consequently, mitigating the noise and biases in audio-based emotion recognition constitutes an important avenue for future research to stabilize multimodal personality evaluation.

\section{Limitation and future Work}

Despite promising results, \textit{EMMR} has several limitations that highlight areas for improvement. First, the framework relies entirely on carefully designed prompts to guide MLLMs, bypassing task-specific fine-tuning. While this design simplifies deployment and preserves the model's zero-shot flexibility, it leaves the MLLMs sub-optimized for the specific nuances of personality assessment. Second, our approach primarily integrates information at the text semantic level. Although transforming non-verbal behaviors into structured textual descriptions facilitates MLLMs reasoning, this prompt-based fusion may inadvertently lose the fine-grained, continuous spatiotemporal dynamics inherent in raw audiovisual signals. Third, the method's effectiveness is heavily dependent on the stability and reliability of the emotion cues extracted from audio and video inputs. As demonstrated in Section 6.3, noisy or misaligned emotion cues can directly propagate errors into the downstream personality assessment.

Building upon these limitations, several research directions emerge for advancing \textit{EMMR}. To address the lack of task-specific optimization, future work could explore parameter-efficient fine-tuning strategies to boost task-specific accuracy while retaining the model's inherent flexibility \cite{11481797}. Furthermore, developing more sophisticated multimodal fusion mechanisms represents a crucial next step. Future research could investigate adaptive fusion architectures that dynamically assign weights to visual, audio, and textual streams based on their contextual reliability\cite{11458684,yang2025enhancing}. Finally, from a theoretical and practical standpoint, future efforts should focus on mitigating inherent biases within MLLMs and improving model calibration to ensure that the resulting personality assessment are consistent, fair, and interpretable.

\section{Conclusion}

In this paper,  we explored the limitations of relying solely on transcribed verbal responses for automated personality assessment in AVIs.To help mitigate potential misinterpretations when verbal content conflicts with non-verbal behaviors, we introduced \textit{EMMR}, a two-stage framework for MLLMs-based personality assessment in AVIs. By extracting emotion cues from visual, audio, and textual modalities and transforming them into structured semantic descriptions, \textit{EMMR} integrates these cues as auxiliary social and behavioral evidence. This approach allows MLLMs to jointly reason over both the candidates' transcribed responses and their emotional expressions.

Experimental indicate that \textit{EMMR} generally improves zero-shot assessment performance across the evaluated HEXACO dimensions. Further analyses suggest that structured emotion semantics effectively complement textual data, aligning with psychological frameworks. Overall, this study provides empirical insights suggesting that fusing explicitly modeled multimodal behavioral evidence with language-based reasoning represents a meaningful direction for future research in personality assessment.

%Bibliography
\bibliographystyle{unsrt}  
\bibliography{references}

@article{lukacik2022into,
  title={Into the void: A conceptual model and research agenda for the design and use of asynchronous video interviews},
  author={Lukacik, Eden-Raye and Bourdage, Joshua S and Roulin, Nicolas},
  journal={Human Resource Management Review},
  volume={32},
  number={1},
  pages={100789},
  year={2022},
  publisher={Elsevier},
  doi={10.1016/j.hrmr.2020.100789}
}

@article{mehta2020recent,
  title={Recent trends in deep learning based personality detection},
  author={Mehta, Yash and Majumder, Navonil and Gelbukh, Alexander and Cambria, Erik},
  journal={Artificial Intelligence Review},
  volume={53},
  number={4},
  pages={2313--2339},
  year={2020},
  publisher={Springer},
  doi={10.1007/s10462-019-09770-z}
}

@article{hickman2022automated,
  title={Automated video interview personality assessments: Reliability, validity, and generalizability investigations},
  author={Hickman, Louis and Bosch, Nigel and Ng, Vincent and Saef, Rachel and Tay, Louis and Woo, Sang Eun},
  journal={Journal of Applied Psychology},
  volume={107},
  number={8},
  pages={1323},
  year={2022},
  publisher={American Psychological Association},
  doi={10.1037/apl0000695}
}

@article{liff2024psychometric,
  title={Psychometric properties of automated video interview competency assessments},
  author={Liff, Josh and Mondragon, Nathan and Gardner, Cari and Hartwell, Christopher J and Bradshaw, Adam},
  journal={Journal of Applied Psychology},
  volume={109},
  number={6},
  pages={921},
  year={2024},
  publisher={American Psychological Association},
  doi={10.1037/apl0001173}
}

@article{achiam2023gpt,
  author       = {Achiam, Josh and others},
  title        = {{GPT-4} Technical Report},
  howpublished = {arXiv:2303.08774},
  month        = mar,
  year         = {2023},
  note         = {[Online]. Available: https://arxiv.org/abs/2303.08774}
}

@misc{bai2025qwen2,
  author       = {Bai, Shuai and others},
  title        = {{Qwen2.5-VL} Technical Report},
  howpublished = {arXiv:2502.13923},
  month        = feb,
  year         = {2025},
  note         = {[Online]. Available: https://arxiv.org/abs/2502.13923}
}

@article{team2024gemma,
  author       = {{Gemma Team} and others},
  title        = {{Gemma 2}: Improving Open Language Models at a Practical Size},
  howpublished = {arXiv:2408.00118},
  month        = aug,
  year         = {2024},
  note         = {[Online]. Available: https://arxiv.org/abs/2408.00118}
}

@article{tan2025prompting,
  title={Prompting-in-a-series: Psychology-informed contents and embeddings for personality recognition with decoder-only models},
  author={Tan, Jing Jie and Kwan, Ban-Hoe and Ng, Danny Wee-Kiat and Hum, Yan-Chai and Mokraoui, Anissa and Lo, Shih-Yu},
  journal={IEEE Transactions on Computational Social Systems},
  year={2025},
  publisher={IEEE},
  doi={10.1109/TCSS.2025.3593323}
}

@article{zhang2024can,
  title={Can large language models assess personality from asynchronous video interviews? A comprehensive evaluation of validity, reliability, fairness, and rating patterns},
  author={Zhang, Tianyi and Koutsoumpis, Antonis and Oostrom, Janneke K and Holtrop, Djurre and Ghassemi, Sina and De Vries, Reinout E},
  journal={IEEE Transactions on Affective Computing},
  volume={15},
  number={3},
  pages={1769--1785},
  year={2024},
  publisher={IEEE},
  doi={10.1109/TAFFC.2024.3374875}
}

@article{zhang2026mixture,
  title={Mixture-of-Expert Large Language Models for text-based Personality Assessment from Asynchronous Video Interviews},
  author={Zhang, Tianyi and Liang, Shan and Zheng, Wenming and Koutsoumpis, Antonis and Oostrom, Janneke K and de Vries, Reinout E},
  journal={IEEE Transactions on Affective Computing},
  year={2026},
  publisher={IEEE},
  doi={ 10.1109/TAFFC.2026.3667846}
}

@ARTICLE{11481797,
  author={Zhang, Tianyi and Hu, Dongsheng and Zhan, Xiu-xiu and Liang, Shan and Zong, Yuan and Li, Yong and Liu, Chuang and Zheng, Wenming},
  journal={IEEE Transactions on Affective Computing}, 
  title={PersonalityLLM: Fine-tuning Large Language Models for Personality Assessment from Asynchronous Video Interviews}, 
  year={2026},
  volume={},
  number={},
  pages={1-15},
  doi={10.1109/TAFFC.2026.3684073}}

@article{breil202113,
  title={13 contributions of nonverbal cues to the accurate judgment of personality traits},
  author={Breil, Simon M and Osterholz, Sarah and Nestler, Steffen and Back, Mitja D},
  journal={The Oxford handbook of accurate personality judgment},
  pages={195--218},
  year={2021}
}

@inbook{Realistic,
author = {Letzring, Tera D. and Colman, Douglas E. and Krzyzaniak, Sheherezade L. and Roberts, Barbara Wood},
publisher = {John Wiley \& Sons, Ltd},
isbn = {9781119547143},
title = {Realistic Accuracy Model},
booktitle = {The Wiley Encyclopedia of Personality and Individual Differences},
chapter = {},
pages = {341-349},
doi = {https://doi.org/10.1002/9781119547143.ch57},
url = {https://onlinelibrary.wiley.com/doi/abs/10.1002/9781119547143.ch57},
year = {2020}
}

@ARTICLE{8999746,
  author={Escalante, Hugo Jair and Kaya, Heysem and Salah, Albert Ali and Escalera, Sergio and Güçlütürk, Yağmur and Güçlü, Umut and Baró, Xavier and Guyon, Isabelle and Junior, Julio C. S. Jacques and Madadi, Meysam and Ayache, Stephane and Viegas, Evelyne and Gürpınar, Furkan and Wicaksana, Achmadnoer Sukma and Liem, Cynthia C. S. and van Gerven, Marcel A. J. and van Lier, Rob},
  journal={IEEE Transactions on Affective Computing}, 
  title={Modeling, Recognizing, and Explaining Apparent Personality From Videos}, 
  year={2022},
  volume={13},
  number={2},
  pages={894-911},
  doi={10.1109/TAFFC.2020.2973984}}

@ARTICLE{9792204,
  author={Jain, Deepak Kumar and Rahate, Anil and Joshi, Gargi and Walambe, Rahee and Kotecha, Ketan},
  journal={IEEE Transactions on Computational Social Systems}, 
  title={Employing Co-Learning to Evaluate the Explainability of Multimodal Sentiment Analysis}, 
  year={2024},
  volume={11},
  number={4},
  pages={4673-4680},
  doi={10.1109/TCSS.2022.3176403}}

@article{yin2024survey,
  title   = {A Survey on Multimodal Large Language Models},
  author  = {Yin, Shukang and Fu, Chaoyou and Zhao, Sirui and Li, Ke and Sun, Xing and Xu, Tong and Chen, Enhong},
  journal = {National Science Review},
  volume  = {11},
  number  = {12},
  pages   = {nwae403},
  year    = {2024},
  month   = dec,
  doi     = {10.1093/nsr/nwae403}
}

@article{suen2024artificial,
  author={Suen, Hung-Yue and Hung, Kuo-En and Liu, Che-Wei and Su, Yu-Sheng and Fan, Han-Chih},
  journal={IEEE Transactions on Computational Social Systems}, 
  title={Artificial Intelligence Can Recognize Whether a Job Applicant Is Selling and/or Lying According to Facial Expressions and Head Movements Much More Correctly Than Human Interviewers}, 
  year={2024},
  volume={11},
  number={5},
  pages={5949-5960},
  doi={10.1109/TCSS.2024.3376732}}

@inproceedings{zhang2025assessing,
  title     = {Assessing Personality Traits and Interview Performance from Asynchronous Video Interviews},
  author    = {Zhang, Tianyi and Qi, Tianhua and Koutsoumpis, Antonios and Zong, Yuan and Zheng, Wenming and Oostrom, Janneke K. and Holtrop, Djurre and Luo, Zhaojie and de Vries, Reinout E.},
  booktitle = {Proceedings of the 33rd ACM International Conference on Multimedia},
  pages     = {13895--13900},
  year      = {2025},
  publisher = {Association for Computing Machinery},
  doi       = {10.1145/3746027.3762016}
}

@ARTICLE{9868797,
  author={Liu, Feng and Wang, Han-Yang and Shen, Si-Yuan and Jia, Xun and Hu, Jing-Yi and Zhang, Jia-Hao and Wang, Xi-Yi and Lei, Ying and Zhou, Ai-Min and Qi, Jia-Yin and Li, Zhi-Bin},
  journal={IEEE Transactions on Computational Social Systems}, 
  title={OPO-FCM: A Computational Affection Based OCC-PAD-OCEAN Federation Cognitive Modeling Approach}, 
  year={2023},
  volume={10},
  number={4},
  pages={1813-1825},
  doi={10.1109/TCSS.2022.3199119}}

@article{marengo2021meta,
  title={A Meta-Analysis on Individual Differences in Primary Emotional Systems and Big Five Personality Traits},
  author={Marengo, Davide and Davis, Kenneth L. and Gradwohl, Gokce Ozkarar and Montag, Christian},
  journal={Scientific Reports},
  volume={11},
  number={1},
  pages={7453},
  year={2021},
  doi={10.1038/s41598-021-84366-8}
}

@article{sacchi2026understanding,
  title={Understanding the Relationship between the Big five personality Traits and the cognitive appraisals Leading to emotions: An integrative narrative review},
  author={Sacchi, Livia and Dan-Glauser, Elise},
  journal={Emotion Review},
  volume={18},
  number={1},
  pages={15--41},
  year={2026},
  publisher={Sage Publications Sage UK: London, England}
}

@ARTICLE{9210819,
  author={Mohammadi, Gelareh and Vuilleumier, Patrik},
  journal={IEEE Transactions on Affective Computing}, 
  title={A Multi-Componential Approach to Emotion Recognition and the Effect of Personality}, 
  year={2022},
  volume={13},
  number={3},
  pages={1127-1139},
  doi={10.1109/TAFFC.2020.3028109}}

@article{zhu2023understanding,
  title={Understanding the Relationships Between Emotion Regulation Strategies and Big Five Personality Traits for Supporting Effective Emotion Regulation Tools/Interventions Design},
  author={Zhu, Ziwen and Qin, Shengfeng and Dodd, Alyson and Conti, Matteo},
  journal={Advanced Design Research},
  volume={1},
  number={1},
  pages={38--49},
  year={2023},
  doi={10.1016/j.ijadr.2023.06.001}
}

@inproceedings{cai2025mdpe,
  author    = {Cai, Can and Liang, Shuang and Liu, Xiao and others},
  title     = {Mdpe: A Multimodal Deception Dataset with Personality and Emotional Characteristics},
  booktitle = {Proceedings of the 33rd ACM International Conference on Multimedia},
  pages     = {12957--12964},
  year      = {2025},
  publisher = {ACM},
  doi       = {10.1145/3746027.3758242},
}

@article{jie2022impact,
  title={Impact of internet usage on consumer impulsive buying behavior of agriculture products: Moderating role of personality traits and emotional intelligence},
  author={Jie, Wei and Poulova, Petra and Haider, Syed Arslan and Sham, Rohana Binti},
  journal={Frontiers in Psychology},
  volume={13},
  pages={951103},
  year={2022},
  publisher={Frontiers Media SA},
  doi={10.3389/fpsyg.2022.951103}
}

@ARTICLE{9969993,
  author={Xu, Shihao and Fang, Jing and Hu, Xiping and Ngai, Edith and Wang, Wei and Guo, Yi and Leung, Victor C. M.},
  journal={IEEE Transactions on Computational Social Systems}, 
  title={Emotion Recognition From Gait Analyses: Current Research and Future Directions}, 
  year={2024},
  volume={11},
  number={1},
  pages={363-377},
  doi={10.1109/TCSS.2022.3223251}}

@article{zhang2024deep,
  title={Deep learning-based multimodal emotion recognition from audio, visual, and text modalities: A systematic review of recent advancements and future prospects},
  author={Zhang, Shiqing and Yang, Yijiao and Chen, Chen and Zhang, Xingnan and Leng, Qingming and Zhao, Xiaoming},
  journal={Expert Systems with Applications},
  volume={237},
  pages={121692},
  year={2024},
  publisher={Elsevier},
  doi={10.1016/j.eswa.2023.121692}
}

@ARTICLE{10537616,
  author={Zhang, Bowen and Ding, Daijun and Huang, Zhichao and Li, Ang and Li, Yangyang and Zhang, Baoquan and Huang, Hu},
  journal={IEEE Transactions on Computational Social Systems}, 
  title={Knowledge-Augmented Interpretable Network for Zero-Shot Stance Detection on Social Media}, 
  year={2025},
  volume={12},
  number={4},
  pages={1773-1784},
  doi={10.1109/TCSS.2024.3388723}}

@article{ashton2007empirical,
  title={Empirical, theoretical, and practical advantages of the HEXACO model of personality structure},
  author={Ashton, Michael C and Lee, Kibeom},
  journal={Personality and Social Psychology Review},
  volume={11},
  number={2},
  pages={150--166},
  year={2007},
  publisher={SAGE Publications Sage CA: Los Angeles, CA},
  doi={10.1177/1088868306294907}
}

@article{krivoshchekov2024passion,
  title={Passion is key: High emotionality in diversity statements promotes organizational attractiveness},
  author={Krivoshchekov, Vladislav and Graf, Sylvie and Sczesny, Sabine},
  journal={British Journal of Social Psychology},
  volume={63},
  number={2},
  pages={544--571},
  year={2024},
  publisher={Wiley Online Library},
  doi={10.1111/bjso.12693}
}

@article{tett2021trait,
  title={Trait activation theory: A review of the literature and applications to five lines of personality dynamics research},
  author={Tett, Robert P and Toich, Margaret J and Ozkum, S Burak},
  journal={Annual Review of Organizational Psychology and Organizational Behavior},
  volume={8},
  number={1},
  pages={199--233},
  year={2021},
  publisher={Annual Reviews},
  doi={10.1146/annurev-orgpsych-012420-062228}
}

@ARTICLE{9003524,
  author={Marouf, Ahmed Al and Hasan, Md. Kamrul and Mahmud, Hasan},
  journal={IEEE Transactions on Computational Social Systems}, 
  title={Comparative Analysis of Feature Selection Algorithms for Computational Personality Prediction From Social Media}, 
  year={2020},
  volume={7},
  number={3},
  pages={587-599},
  doi={10.1109/TCSS.2020.2966910}}

@ARTICLE{9531972,
  author={K. N., Pavan Kumar and Gavrilova, Marina L.},
  journal={IEEE Transactions on Computational Social Systems}, 
  title={Latent Personality Traits Assessment From Social Network Activity Using Contextual Language Embedding}, 
  year={2022},
  volume={9},
  number={2},
  pages={638-649},
  doi={10.1109/TCSS.2021.3108810}}

@ARTICLE{9760455,
  author={Sun, Xiangguo and Liu, Bo and Ai, Liya and Liu, Danni and Meng, Qing and Cao, Jiuxin},
  journal={IEEE Transactions on Computational Social Systems}, 
  title={In Your Eyes: Modality Disentangling for Personality Analysis in Short Video}, 
  year={2023},
  volume={10},
  number={3},
  pages={982-993},
  doi={10.1109/TCSS.2022.3161708}}

@inproceedings{yang2025enhancing,
  title={Enhancing Multimodal Personality Assessment with LLM-Augmented Hierarchical Fusion},
  author={Yang, Longjiang and Yu, Cong and Huang, Chenxi and Zhang, Fengyu and Liu, Ran and Wen, Zhuofan and Chen, Shun and Yao, Hailiang and Liu, Bin and Lian, Zheng and others},
  booktitle={Proceedings of the 33rd ACM International Conference on Multimedia},
  pages={13917--13923},
  year={2025}
}

@article{kassab2026multi,
  title={Multi-Modal Method for Candidate Interview Assessment Based on Computer Vision and Large Language Models},
  author={Kassab, Kenan and Kashevnik, Alexey and Shoshina, Irina},
  journal={Big Data and Cognitive Computing},
  volume={10},
  number={4},
  pages={106},
  year={2026},
  publisher={MDPI}
}

@ARTICLE{11194051,
  author={Feng, Mengling and Cambria, Erik and Lin, Qika and Shi, Kaize and Li, Weiping},
  journal={IEEE Transactions on Computational Social Systems}, 
  title={Guest Editorial: Multimodal Representation and Reasoning for Social Computing}, 
  year={2025},
  volume={12},
  number={5},
  pages={3565-3568},
  doi={10.1109/TCSS.2025.3606567}}

@ARTICLE{10932863,
  author={Wang, Yaowei and Lin, Zulong and Teng, Yan and Cheng, Yuqi and Jiang, Hongyan and Yang, Yun},
  journal={IEEE Transactions on Computational Social Systems}, 
  title={SIMMA: Multimodal Automatic Depression Detection via Spatiotemporal Ensemble and Cross-Modal Alignment}, 
  year={2025},
  volume={12},
  number={5},
  pages={3548-3564},
  doi={10.1109/TCSS.2025.3542986}}

@article{ashton2004six,
  title={A six-factor structure of personality-descriptive adjectives: solutions from psycholexical studies in seven languages},
  author={Ashton, Michael C and Lee, Kibeom and Perugini, Marco and Szarota, Piotr and De Vries, Reinout E and Di Blas, Lisa and Boies, Kathleen and De Raad, Boele},
  journal={Journal of Personality and Social Psychology},
  volume={86},
  number={2},
  pages={356},
  year={2004},
  publisher={American Psychological Association},
  doi={10.1037/0022-3514.86.2.356}
}

@inproceedings{ponce2016chalearn,
  title={Chalearn lap 2016: First round challenge on first impressions-dataset and results},
  author={Ponce-L{\'o}pez, V{\'\i}ctor and Chen, Baiyu and Oliu, Marc and Corneanu, Ciprian and Clap{\'e}s, Albert and Guyon, Isabelle and Bar{\'o}, Xavier and Escalante, Hugo Jair and Escalera, Sergio},
  booktitle={European Conference on Computer Vision},
  pages={400--418},
  year={2016},
  organization={Springer},
  doi={10.1007/978-3-319-49409-8_32}
}

@inproceedings{sun2024revealing,
  title={Revealing personality traits: A new benchmark dataset for explainable personality recognition on dialogues},
  author={Sun, Lei and Zhao, Jinming and Jin, Qin},
  booktitle={Proceedings of the 2024 Conference on Empirical Methods in Natural Language Processing},
  pages={19988--20002},
  year={2024}
}

@misc{hui2024qwen2,
  author       = {Hui, Binyuan and others},
  title        = {{Qwen2.5-Coder} Technical Report},
  howpublished = {arXiv:2409.12186},
  month        = sep,
  year         = {2024},
  note         = {[Online]. Available: https://arxiv.org/abs/2409.12186}
}

@article{wang2024qwen2,
  author       = {Wang, Peng and others},
  title        = {{Qwen2-VL}: Enhancing Vision-Language Model's Perception of the World at Any Resolution},
  howpublished = {arXiv:2409.12191},
  month        = sep,
  year         = {2024},
  note         = {[Online]. Available: https://arxiv.org/abs/2409.12191}
}

@article{team2025kimi,
  author       = {{Kimi Team} and others},
  title        = {{Kimi-VL} Technical Report},
  howpublished = {arXiv:2504.07491},
  month        = apr,
  year         = {2025},
  note         = {[Online]. Available: https://arxiv.org/abs/2504.07491}
}

@article{ding2025kimi,
  author       = {Ding, Ding and others},
  title        = {{Kimi-Audio} Technical Report},
  howpublished = {arXiv:2504.18425},
  month        = apr,
  year         = {2025},
  note         = {[Online]. Available: https://arxiv.org/abs/2504.18425}
}

@article{wei2017deep,
  title={Deep bimodal regression of apparent personality traits from short video sequences},
  author={Wei, Xiu-Shen and Zhang, Chen-Lin and Zhang, Hao and Wu, Jianxin},
  journal={IEEE Transactions on Affective Computing},
  volume={9},
  number={3},
  pages={303--315},
  year={2017},
  publisher={IEEE},
  doi={10.1109/TAFFC.2017.2762299}
}

@inproceedings{girdhar2019video,
  title={Video action transformer network},
  author={Girdhar, Rohit and Carreira, Joao and Doersch, Carl and Zisserman, Andrew},
  booktitle={Proceedings of the IEEE/CVF conference on computer vision and pattern recognition},
  pages={244--253},
  year={2019},
doi={}
}

@inproceedings{liu2021swin,
  title={Swin transformer: Hierarchical vision transformer using shifted windows},
  author={Liu, Ze and Lin, Yutong and Cao, Yue and Hu, Han and Wei, Yixuan and Zhang, Zheng and Lin, Stephen and Guo, Baining},
  booktitle={Proceedings of the IEEE/CVF international conference on computer vision},
  pages={10012--10022},
  year={2021},
doi={}
}

@article{duijndam2024behavior,
  title   = {Behavior and Emotion Regulation of Socially Inhibited Individuals in Specific Uncomfortable Situations},
  author  = {Duijndam, Stefanie and others},
  journal = {International Journal of Clinical and Health Psychology},
  volume  = {24},
  number  = {4},
  pages   = {100502},
  year    = {2024},
  doi     = {10.1016/j.ijchp.2024.100502}
}

@ARTICLE{10542112,
  author={Zhang, Zixing and Peng, Liyizhe and Pang, Tao and Han, Jing and Zhao, Huan and Schuller, Björn W.},
  journal={IEEE Transactions on Computational Social Systems}, 
  title={Refashioning Emotion Recognition Modeling: The Advent of Generalized Large Models}, 
  year={2024},
  volume={11},
  number={5},
  pages={6690-6704},
  doi={10.1109/TCSS.2024.3396345}}

@article{koutsoumpis2024beyond,
  title={Beyond traditional interviews: Psychometric analysis of asynchronous video interviews for personality and interview performance evaluation using machine learning},
  author={Koutsoumpis, Antonis and Ghassemi, Sina and Oostrom, Janneke K and Holtrop, Djurre and van Breda, Ward and Zhang, Tianyi and de Vries, Reinout E},
  journal={Computers in Human Behavior},
  volume={154},
  pages={108128},
  year={2024},
  publisher={Elsevier},
  doi={10.1016/j.chb.2023.108128}
}

@misc{dataset2025,
  author       = {Koutsoumpis, Antonios and Zhang, Tianyi and Oostrom, Janneke K. and Holtrop, Djurre and de Vries, Reinout E.},
  title        = {Open Dataset {AVI-6}: Annotated Asynchronous Video Interviews},
  howpublished = {Open Science Framework},
  year         = {2025},
  doi          = {10.17605/OSF.IO/XTUYQ},
  note         = {[Online]. Available: https://doi.org/10.17605/OSF.IO/XTUYQ}
}

@ARTICLE{11458684,
  author={Wang, Yusong and Li, Dongyuan and Funakoshi, Kotaro and Okumura, Manabu},
  journal={IEEE Access}, 
  title={E²MP: Emotion-Enhanced Modality Fusion With Contrastive Learning for Multi-Modal Personality Traits Recognition}, 
  year={2026},
  volume={14},
  number={},
  pages={50516-50536},
  doi={10.1109/ACCESS.2026.3679356}}

\end{document}